\documentclass[11pt,prd,twocolumn]{revtex4-1}
\usepackage{graphicx, amsmath, amssymb, color}
\def\phic{\phi_r}
\def\aap{A\&A}
\def\aj{Astron. J.}
\def\mnras{MNRAS}
\def\apjs{Astrophys. J. Suppl.}
\newcommand{\beq}{\begin{equation}}
\newcommand{\eeq}{\end{equation}}

\begin{document}
\title{Revisiting Ryskin's Model of Cosmic Acceleration}
\author{Zhiqi Huang}
\affiliation{School of Physics and Astronomy, Sun Yat-sen University, 2 Daxue Road, Tangjia, Zhuhai, China}
\email{huangzhq25@mail.sysu.edu.cn}
\author{Han Gao}
\affiliation{School of Physics and Astronomy, Sun Yat-sen University, 2 Daxue Road, Tangjia, Zhuhai, China}
\author{Haoting Xu}
\affiliation{School of Physics and Astronomy, Sun Yat-sen University, 2 Daxue Road, Tangjia, Zhuhai, China}
\date{\today}
\begin{abstract}
  Cosmic backreaction as an additional source of the expansion of the universe has been a debate topic since the discovery of cosmic acceleration. The major concern is whether the self interaction of small-scale nonlinear structures would source gravity on very large scales. Gregory Ryskin argued against the additional inclusion of gravitational interaction energy of astronomical objects, whose masses are mostly inferred from gravitational effects and hence should already contain all sources with long-range gravity forces. Ryskin proposed that the backreaction contribution to the energy momentum tensor comes instead from the rest of the universe beyond the observable patch. Ryskin's model solves the fine-tuning problem and is in good agreement with the Hubble diagram of Type Ia supernovae. In this article we revisit Ryskin's model and show that it is {\it inconsistent} with at least one of the following statements: (i) the universe is matter-dominated at low redshift ($z\lesssim 2$); (ii) the universe is radiation-dominated at sufficiently high redshift; (iii) matter density fluctuations are tiny ($\lesssim 10^{-4}$) at the recombination epoch. 
 \end{abstract}

\maketitle

\section{Introduction}

Observations of cosmic microwave background (CMB) radiation have revealed that the primordial universe is nearly homogeneous (density fluctuations $\lesssim 10^{-4}$) on a wide range of cosmological scales from a few Mpc to tens of Gpc~\cite{COBE, WMAPParam09, PlanckParam15, PlanckParam18}. At low redshift ($z\lesssim 2$) we observe in contrast a matter dominated universe with hierarchical structures, from galaxies, groups of galaxies, clusters and superclusters to the large-scale cosmic web with filaments and voids. The growth of cosmic structures shows the gravitational instability caused by the attractive nature of gravity. On the other hand,  the accelerated expansion of the late-time universe ($z\lesssim 0.5$), first inferred from the type Ia supernovae light curves~\citep{Riess98, Perlmutter99} and later supported by many other evidence~\cite{PlanckParam15, PlanckParam18, BAO-SDSS-DR12-LOWZ, BAO-SDSS-DR12-CMASS,DESParams18, DESWL18, Abbott17}, implies a repulsive force on very large scales. 

In the concordance picture of modern cosmology, the hierarchical structures and the late-time accelerated expansion of the universe can be explained by inclusion of cold dark matter (CDM) and dark energy, respectively. On large scales the coarse-grained universe is described by perturbed Einstein's equations, whereas numerical simulations completes the story on smaller scales. Among many theoretical constructions the simplest version, known as the $\Lambda$CDM model, where dark energy is interpreted as Einstein's cosmological constant $\Lambda$, is so far in good agreement with most observational data, and thus is favored by the principle of Occam's razor. There are a few instances of claimed observational evidence against $\Lambda$CDM model~\cite{Coldspot1, Coldspot2, Bennett03, Bennett11, Ade13, Ade15, Cautun16, Riess18, Riess19, Buchert15}, most of which involve modeling of complex astrophysics and are yet under debate.

Despite its great success, $\Lambda$CDM model is not conclusively an end mark of cosmology. If the fine-tuning nature of cosmological constant~\cite{Weinberg89, Weinberg00, Mannheim17} only philosophically disturbs cosmologists, the lack of a solid proof that the coarse graining approach is applicable to Einstein's equations, which are nonlinear and do not commute with Fourier filtering, is probably a more serious concern.

The Einstein's equations for coarse-grained metric $g_{\mu\nu}$ and coarse-grained energy-momentum tensor $T_{\mu\nu}$ are written in natural units ($c=\hbar=G=1$) as~\cite{Buchert00, BRreview}
\begin{equation}
G_{\mu\nu} = 8\pi \left(T_{\mu\nu}+B_{\mu\nu}\right), \label{eq:B}
\end{equation}
where $G_{\mu\nu}$ is Einstein tensor for $g_{\mu\nu}$ and $B_{\mu\nu}$ accounts for the difference between $G_{\mu\nu}$ and coarse-grained Einstein tensor. The standard interpretation is that $B_{\mu\nu}$ arises from small-scale gravity self-interaction energy that backreacts to large scales~\cite{Buchert00}. This interpretation was criticized by Gregory Ryskin, who argued that the observed $T_{\mu\nu}$ is inferred from gravitational effects, and thus already contains all sources of long-range gravity forces, including the small-scale gravity self-interaction energy~\cite{Ryskin15}. Ryskin proposed instead a contribution from the rest of the universe beyond the observable patch 
\beq
B_{\mu \nu} = \rho_m \phi_c g_{\mu \nu}, \label{eq:phicm}
\eeq
where $\rho_m$ is the matter density and $\phi_c$ is a constant. By some simple reasoning, Ryskin identified $\phi_c= -3$ for a universe that is spatially flat and dominated by non-relativistic matter.

Eq.~\eqref{eq:phicm} can be cast into a perfect-fluid form with an effective energy density 
\begin{equation}
  \rho = \rho_m\left(1-\phi_c\right) = 4\rho_m,  \label{eq:totrhom}
\end{equation}
and an effective pressure
\begin{equation}
  p = \phi_c\rho_m = -3\rho_m.  \label{eq:totpm}
\end{equation}
The conservation of energy, or in Ryskin's terminology, the first law of thermodynamics leads to
\begin{equation}
  \rho_m \propto a^{-3/4}, \label{eq:rhom}
\end{equation}
and
\begin{equation}
  a\propto t^{8/3}, \label{eq:atm}
\end{equation}
where $a$ is the scale factor and $t$ is the cosmological time.

Eq.~\eqref{eq:atm} seems to be radical as it predicts cosmic acceleration for the entire matter-dominated era. However, because the usual assumption $\rho_m\propto a^{-3}$ no longer holds in Ryskin's model, caution needs to be taken for comparisons with the observational data. It has been shown that Ryskin's model is a good fit to the Hubble diagram from Type Ia supernovae data~\cite{Ryskin15}.

A natural question then arises whether Ryskin's model can explain the formation of the large-scale structures of the universe. By solving the matter density perturbation equation
\begin{equation}
  \ddot\delta + 2H\dot\delta = 4\pi \rho_m\delta, \label{eq:growth}
\end{equation}
where a dot denotes derivative with respect to $t$ and $H=\dot a/a$ is the Hubble parameter, Ryskin obtained the linear growth of matter density fluctuations
\begin{equation}
  \delta \propto t^{0.5465}\propto a^{0.2049}. \label{eq:deltam}
\end{equation}
Recalling in standard paradigm $\delta\propto a$ in matter-dominated era, the growth of structure seems to be too slow in Ryskin's model. Unfortunately, due to an unknown bias factor between galaxies and cold dark matter, we do not directly observe $\delta$ from galaxy clustering. The indirect constraints on $\delta$ are mainly from redshift-space distortion that is model-dependent. It is yet unclear how to self-consistently derive redshift-space distortion effect from Ryskin's model. In Ref.~\cite{Ryskin15} Ryskin made some simple estimations and claimed that Eq.~\eqref{eq:deltam} may be roughly consistent with the observed clustering of galaxies.

Ryskin's discussion, however, may have missed a potentially more serious problem with Eq.~\eqref{eq:deltam}. If recombination is at redshift $z_{\rm rec}\sim 1000$ and the universe has been matter-dominated since then, today's matter density fluctuations would remain tiny $\delta_0\lesssim 10^{-4}\times 1000^{0.2049}\lesssim 10^{-3}$, apparently in contradiction with the observed large-scale structures at low redshift.

Again caution needs to be taken for the above arguments. In Ryskin's model neither $z_{\rm rec}\sim 10^3$ nor matter domination at $z<z_{\rm rec}$ is guaranteed. Even the radiation-dominated era and the recombination epoch are not discussed in details, and not guaranteed to exist in Ryskin's original work~\cite{Ryskin15}. We will nevertheless work with the basic picture that a radiation-dominated and  nearly homogeneous universe evolves into a matter-dominated universe with significant density fluctuations. In this context, we will show that Eq.~\eqref{eq:deltam} and hence Ryskin's model is indeed {\it inconsistent} with observations.

\section{Recombination in Ryskin's Model}

To make our discussion as general as possible, we will allow early dark energy in the radiation-dominated era, too. Following Ryskin's philosophy~\cite{Ryskin15}, the total energy density and pressure for a radiation-dominated universe are
\begin{eqnarray}
  \rho &=& \left(1-\phic\right)\rho_r,  \label{eq:totrho} \\
  p &=& \left(\frac{1}{3}+\phic\right)\rho_r,  \label{eq:totp}
\end{eqnarray}
where $\rho_r$ is the energy density of radiation, and $\phic$, as an anolog to $\phi_c$, is a constant. A negative $\phic$ corresponds to a positive dark energy density, whereas the $\phic=0$ limit corresponds to the standard radiation component without early dark energy. Ryskin derived $\phi_c=-3$ for a matter-dominated universe. Unfortunately, the non-relativistic limit used in Ryskin's derivation no longer works for radiation and there is no obvious way to determine the value of $\phic$. Thus, we will leave $\phic$ as a free parameter.

The first law of thermodynamics is
\begin{equation}
  d(\rho a^3) = - p d (a^3), \label{eq:1stlaw}
\end{equation}
which together with Eq.~\eqref{eq:totrho} and Eq.~\eqref{eq:totp} implies
\begin{equation}
  \rho_r \propto a^{-\frac{4}{1-\phic}}. \label{eq:rhorpower}
\end{equation}
It follows from Eq.~\eqref{eq:rhom} and Eq.~\eqref{eq:rhorpower} that the redshift of matter-radiation equality is
\begin{equation}
z_{\rm eq} = \left(5938h^2\right)^{\frac{4(1-\phic)}{13+3\phic}}-1,
\end{equation}
where we have used $\Omega_m=0.25$ for Ryskin's model and assumed three species of light neutrinos. The universe at low redshift ($z\lesssim 2$) is evidently matter-dominated. The existence of an $2<z_{\rm eq}<\infty$ thus leads to
\begin{equation}
 -4.3 < \phic < 0.5.
\end{equation}

The number of photons is not conserved for a nonzero $\phic$. The energy of a photon is $\propto (1+z)$. Thus, the number of photons evolves as
\begin{equation} 
n_\gamma \propto \frac{\rho_r}{1+z} \propto a^{- \frac{3 + \phic}{1-\phic}}. \label{eq:ngamma}
\end{equation}

At the moment of recombination, the ratio between the number of baryons and the number of photons $ \frac{n_B}{n_\gamma}$  is about the Boltzmann factor for the ionization of a hydrogen atom
\begin{equation}
\frac{n_B}{n_\gamma} = e^{-\frac{E}{\eta k T}},  \label{eq:bgrat1}
\end{equation}
where $E=13.6\mathrm{eV}$, $T =2.73(1+z) \mathrm{K}$ and $k$ is the Boltzmann constant. The numeric factor $\eta\sim O(1)$ captures the detailed physics of recombination. For the standard cosmology $\eta \approx 2.5$. We will demonstrate that our conclusion does not depend on the detailed value of $\eta$.

The complexity arises from that in Ryskin's model the ratio $n_B/n_\gamma$ is redshift-dependent. Scaling the ratio from today's measured value of $n_\gamma$, we have
\begin{equation}
  \frac{n_B}{n_\gamma} = 2.75\times 10^{-8} \Omega_b h^2 (1+z)^{\frac{3}{4} - \frac{3 + \phic}{1-\phic}}, \label{eq:bgrat2}
\end{equation}
where $\Omega_bh^2$ is the baryon density parameter. In standard cosmology it is measured to be $\Omega_bh^2\approx 0.022$. In Ryskin's model and in a more radical scenario without cold dark matter, we may have $\Omega_bh^2\sim  0.1$. We will show that our result is not sensitive to the value of $\Omega_b h^2$, neither.

Substituting Eq.~\eqref{eq:bgrat2} into Eq.~\eqref{eq:bgrat1}, we obtain a recombination redshift $z_{\rm rec}$ that depends on $\phic$, $\Omega_b h^2$, and $\eta$.
\begin{figure}
  \includegraphics[width=0.48\textwidth]{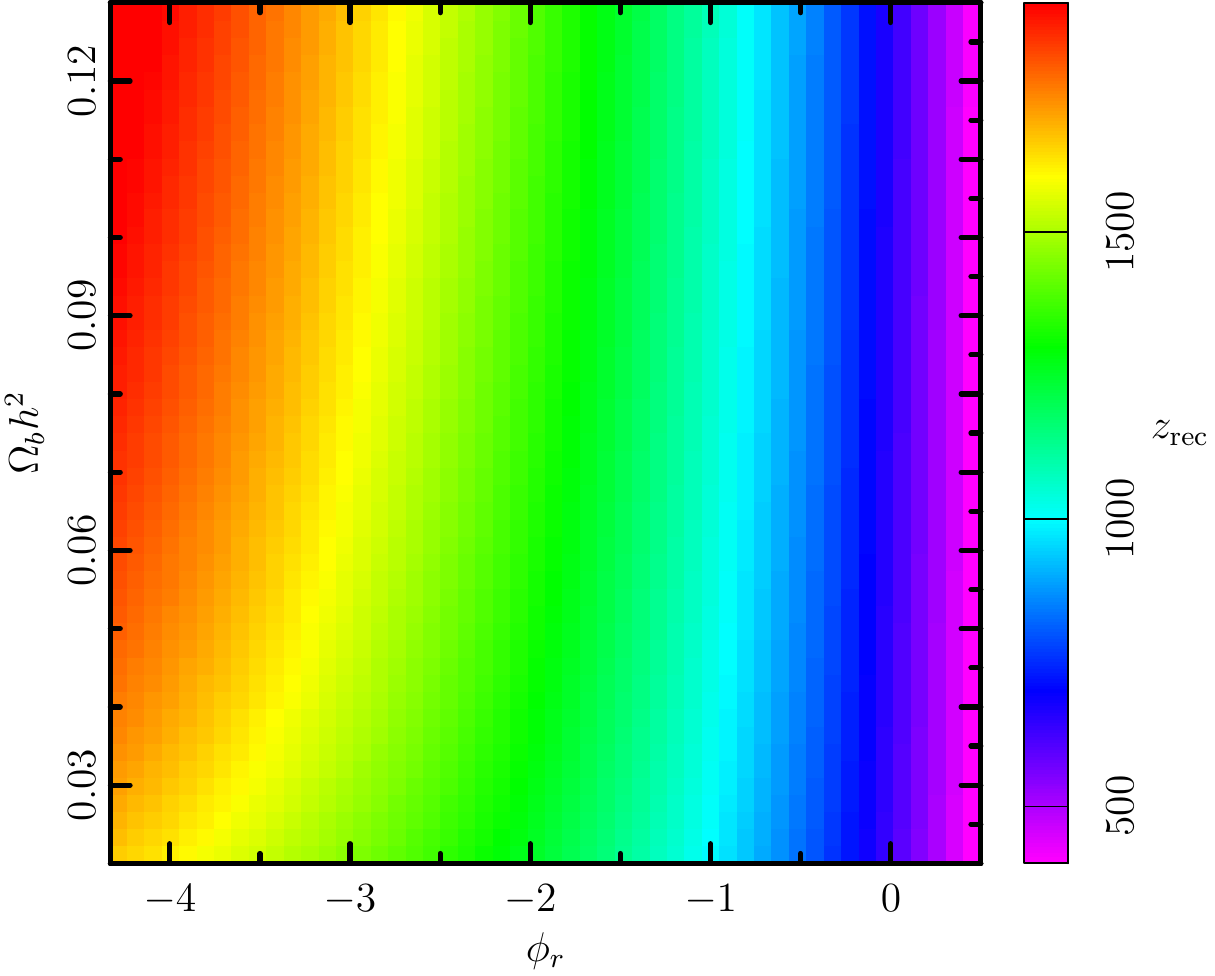}
  \caption{The recombination redshift in Ryskin's model for fixed $\eta=2.5$.   \label{fig:zrec}}
\end{figure}
In Fig~\ref{fig:zrec} we show that $z_{\rm rec}$ is $\sim O(10^3)$ for fixed $\eta = 2.5$. The scaling of $1+z_{\rm rec}$ with $\eta$ is linear and thus variation of $\eta\sim O(1)$ will not change the qualitative conclusion.

\section{Growth of Structures}

Before carrying out an exact calculation of $\delta(t)$, we may intuitively expect free streaming of radiation to suppress the growth of matter density contrast. If we approximate $\delta$ as a constant during radiation-dominated era, the total growth of $\delta$ from recombination epoch to today is roughly
\begin{equation}
  \frac{\delta_0}{\delta_{\rm rec}} \sim \left[1 + \min\left(z_{\rm rec}, z_{\rm eq}\right)\right]^{0.2049}. \label{eq:approx}
\end{equation}

To obtain a more rigorous result, we consider Eq.~\eqref{eq:growth} in the presence of both radiation and matter. The Hubble parameter $H$ in Eq.~\eqref{eq:growth} is derived from the total energy density, which is the sum of Eq.~\eqref{eq:totrhom} and Eq.~\eqref{eq:totrho}. We start the evolution of $\delta$ at $a\rightarrow 0^+$, so that the decaying mode will be suppressed and an initial growing mode at recombination will be set automatically. In all calculations we use a fixed $H_0=70\mathrm{km\,s^{-1}Mpc^{-1}}$. Small variations of $H_0$ do not lead to any visible differences. Thus, hereafter we do not discuss the impact of $H_0$ uncertainty.

\begin{figure}
  \includegraphics[width=0.48\textwidth]{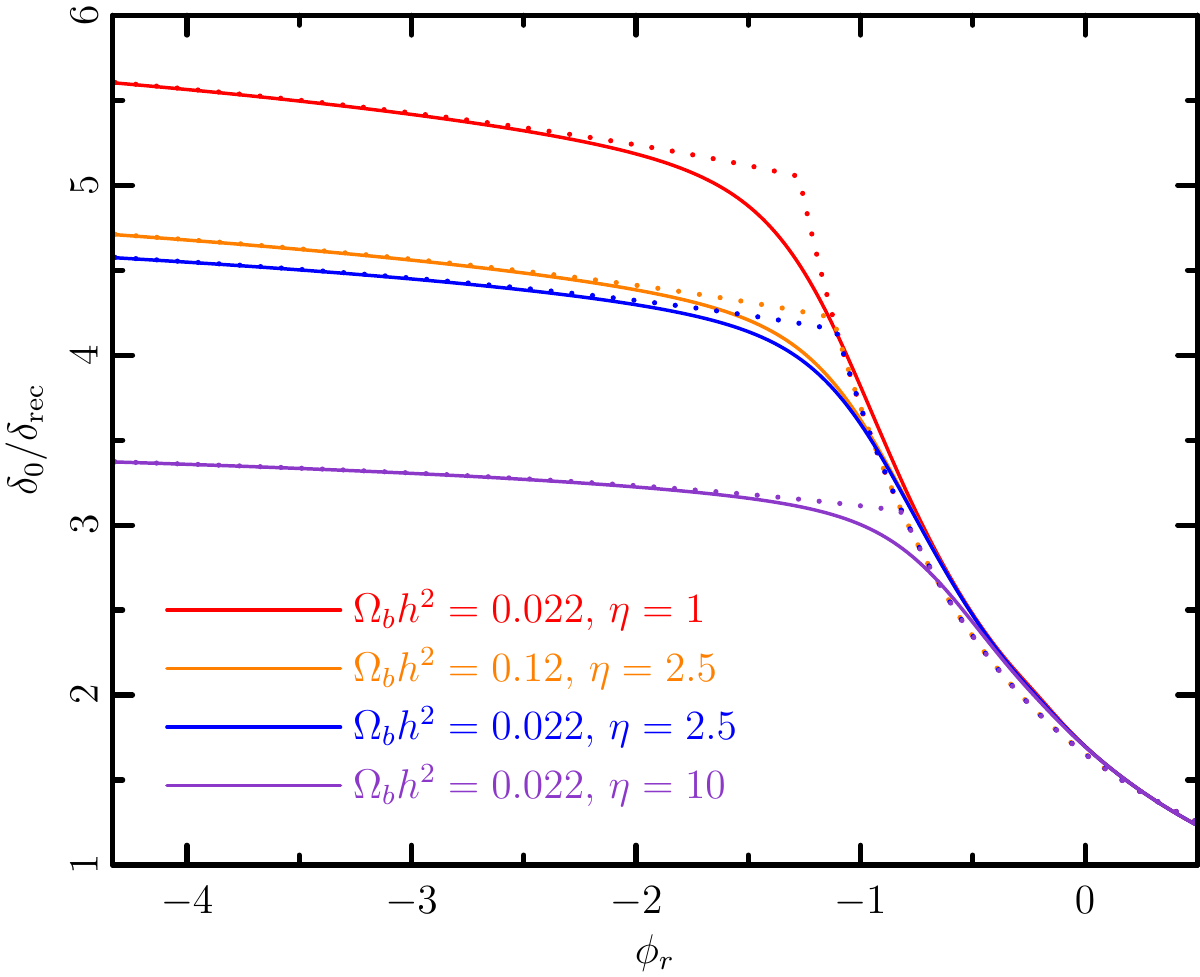}
  \caption{The total growth of matter density contrast from recombination epoch to today. The dotted lines are approximation given by Eq.~\eqref{eq:approx}.   \label{fig:delta}}
\end{figure}

In Fig~\ref{fig:delta} we show the exact solutions of Eq.~\eqref{eq:growth} for a few representative values of $\Omega_bh^2$ and $\eta$. Eq.~\eqref{eq:approx} turns out to be a very good approximation in all cases. The result $\delta_0/\delta_{\rm rec}\sim$ a few is, being order of unity, much too small, and is in contradiction with the much larger ratio of density perturbations today and CMB anisotropies generated at $z\sim 1000$.

\section{Discussion and Conclusions}

Ryskin's model of cosmic acceleration, despite being elegant, fails to explain the observed significant growth of cosmic structures from the recombination epoch to today. Our results are very robust and are insensitive to the details of recombination, the baryon abundance, and how the early dark energy scales with radiation component. Nevertheless, there may still be some room to save the model, for instance, by proposing a radically different form of early dark energy during the radiation-dominated era. Study along this direction is beyond the scope of this paper. We look forward to see if the model can be revised to agree with basic properties of the observed universe.

\end{document}